\documentclass[12pt,a4paper]{article}
\usepackage{graphicx}
\usepackage{times}
\title{\bf Cyg OB2 \#5: When three stars are just not enough.}
\author{M. Kennedy$^1$, S.M. Dougherty$^{2,3}$, P.M. Williams$^4$ and A. Fink$^2$\\
\vspace{1cm}\\
\normalsize $^1$ University of Victoria, BC, Canada\\ 
\normalsize $^2$ NRC-HIA DRAO, Penticton, BC, Canada\\
\normalsize $^3$ Institute for Space Imaging Science, University of Calgary, AB, Canada \\
\normalsize $^4$ Institute for Astronomy, Royal Observatory, Blackford Hill, Edinburgh, Scotland}
\date{\mbox{}}
\begin{document}
\maketitle
\pagestyle{empty}
%
%
\def\bull{\vrule height .9ex width .8ex depth -.1ex}
\makeatletter
\def\ps@plain{\let\@mkboth\gobbletwo
\def\@oddhead{}\def\@oddfoot{\hfil\tiny\bull\quad
``The multi-wavelength view of hot, massive stars''; 39$^{\rm th}$ Li\`ege Int.\ Astroph.\ Coll., 12-16 July 2010 \quad\bull}%
\def\@evenhead{}\let\@evenfoot\@oddfoot}
\makeatother
%
%
\def\beginrefer{\section*{References}%
\begin{quotation}\mbox{}\par}
\def\refer#1\par{{\setlength{\parindent}{-\leftmargin}\indent#1\par}}
\def\endrefer{\end{quotation}}
%
%
{\noindent\small{\bf Abstract:} Archival observations from the Very
Large Array (VLA) at frequencies between 1.4 GHz and 43 GHz of the
6.6-day O6.5-7+O5.5-6 binary {Cyg OB2 \#5} over 20 years are
re-examined. The aim is to determine the location and character of its
known variable radio emission.  The radio emission consists of a
primary component associated with the binary, and a non-thermal source
(NE), $0.8^{\prime\prime}$ to the NE.  This work reveals that NE shows
no evidence of variation demonstrating that the variable
emission arises in the primary component. With NE constant,
the radio flux from the primary can now be well determined for the
first time, especially in observations that do not resolve both
the primary and NE components. The variable radio emission from the
primary has a period of $6.7\pm0.3$~years which is described
by a simple model of a non-thermal source orbiting within the stellar
wind envelope of the binary. Such a model implies the presence of a
third, unresolved stellar companion (Star C) orbiting the 6.6-day
binary with a period of 6.7 years. The non-thermal emission
arises from either a WCR between Star C and the binary system, or
possibly from Star C directly.  Examination of radial velocity
observations suggests reflex motion of the binary due to Star C, for
which a mass of $23^{+22}_{-14}$~M$_{\odot}$ is deduced.  Together
with the star associated with NE, this implies that {Cyg OB2 \#5} is a
quadruple system.

%
%
\section{Introduction}
{Cyg OB2 \#5} ({V729 Cyg}, {BD $+40\,4220$}) is an eclipsing binary
system consisting of two O-type supergiants orbiting in a 6.6-day
period (Hall, 1974; Leung \& Schneider 1978; Rauw et al. 1999). This
system is one of several luminous O-star systems in the Cyg OB2
association that shows evidence of variable radio emission (Persi et
al. 1983, 1990; Bieging et al. 1989). The radio emission has two
states: a low-flux state of $\sim2$~mJy at 4.8 GHz where the spectral
index is consistent with thermal emission from a stellar wind, and a
high-flux state of $\sim8$~mJy at 4.8 GHz, where the spectral index is
flatter than in the low state.  The variations appear to have a
$\sim7$-year period (Miralles et al. 1994) and have been attributed to
variable non-thermal emission from an expanding plasmon arising in the
binary (Bieging et al. 1989; Persi et al. 1990; Miralles et al.
1994).

Observations by Abbott et al. (1981) with the VLA revealed a primary
component associated with the binary and a second component (hereafter
NE) $\sim0.8^{\prime\prime}$ to the NE of the primary radio
source. NE appears non-thermal and Contreras et al. (1997) argue it is
the result of a wind-collision region (WCR) between the stellar wind
of the binary and that of a B0-2 V star, $0.9^{\prime\prime}$ from the
binary to the NE (Contreras et al. 1997).

The previous analyses of the radio emission from {Cyg OB2 \#5} are
based on observations from the Very Large Array (VLA), obtained in all
the different configurations of the array.  Hence, the observations
may or may not resolve the emission from both the primary and NE
components. In this work, all VLA archive radio observations of {Cyg
OB2 \#5} are re-examined to produce a consistently calibrated data
set. This analysis accounts for changes in the resolution of the
observations and presents a consistent treatment of the emission from
both the primary and NE sources in each observation.  For the first
time, the nature and evolution of the emission of both the primary and
NE are determined. Kennedy et al. (2010) provide a more complete
description of the work highlighted here.

\section{Observations}

A total of 50 VLA observations of {Cyg OB2 \#5} obtained between 1983
and 2003 were extracted from the NRAO archive, re-calibrated and
analysed.  Examples of the resulting deconvolved images at 8.4~GHz are
shown in Fig.~\ref{fig:images}.

\begin{figure}[h]
\includegraphics[width=8cm]{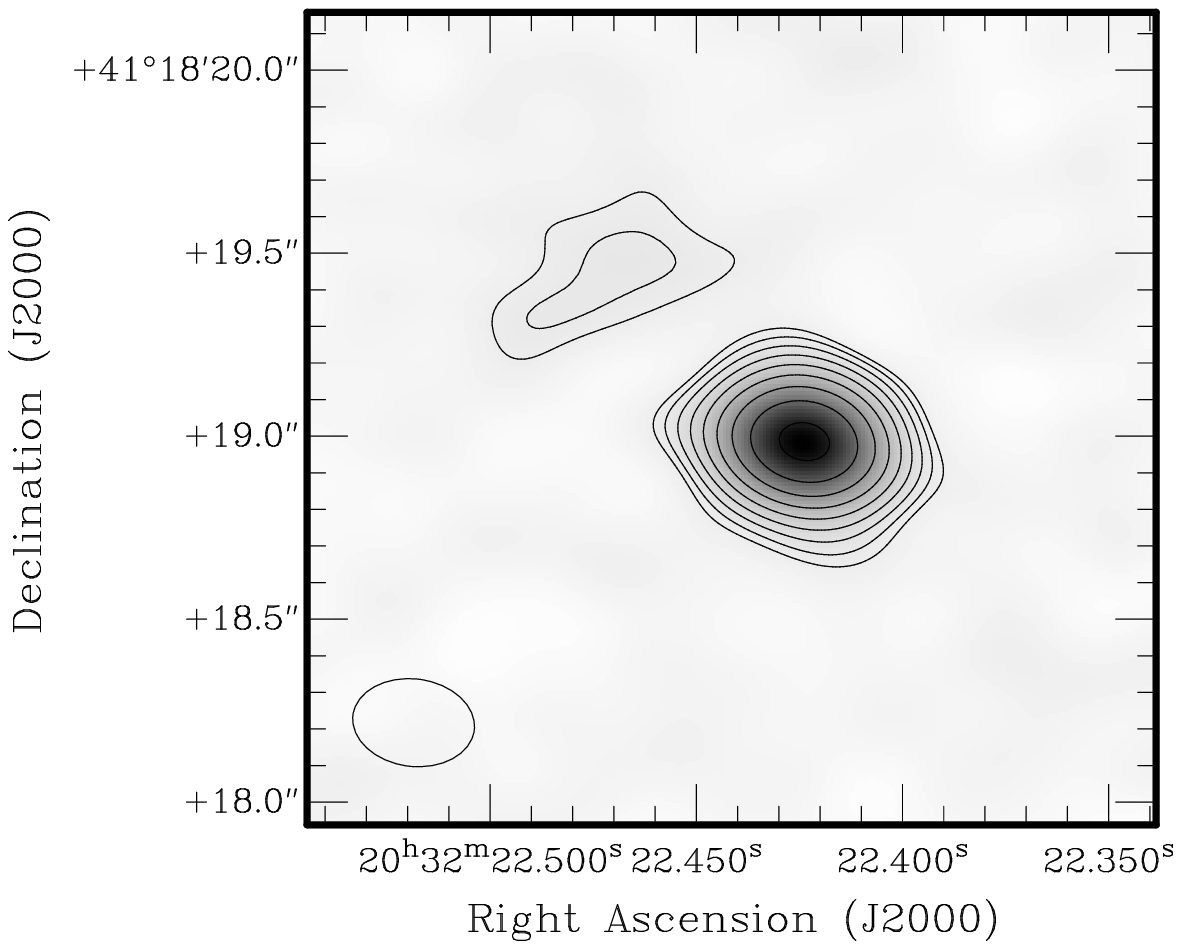}
\includegraphics[width=8cm]{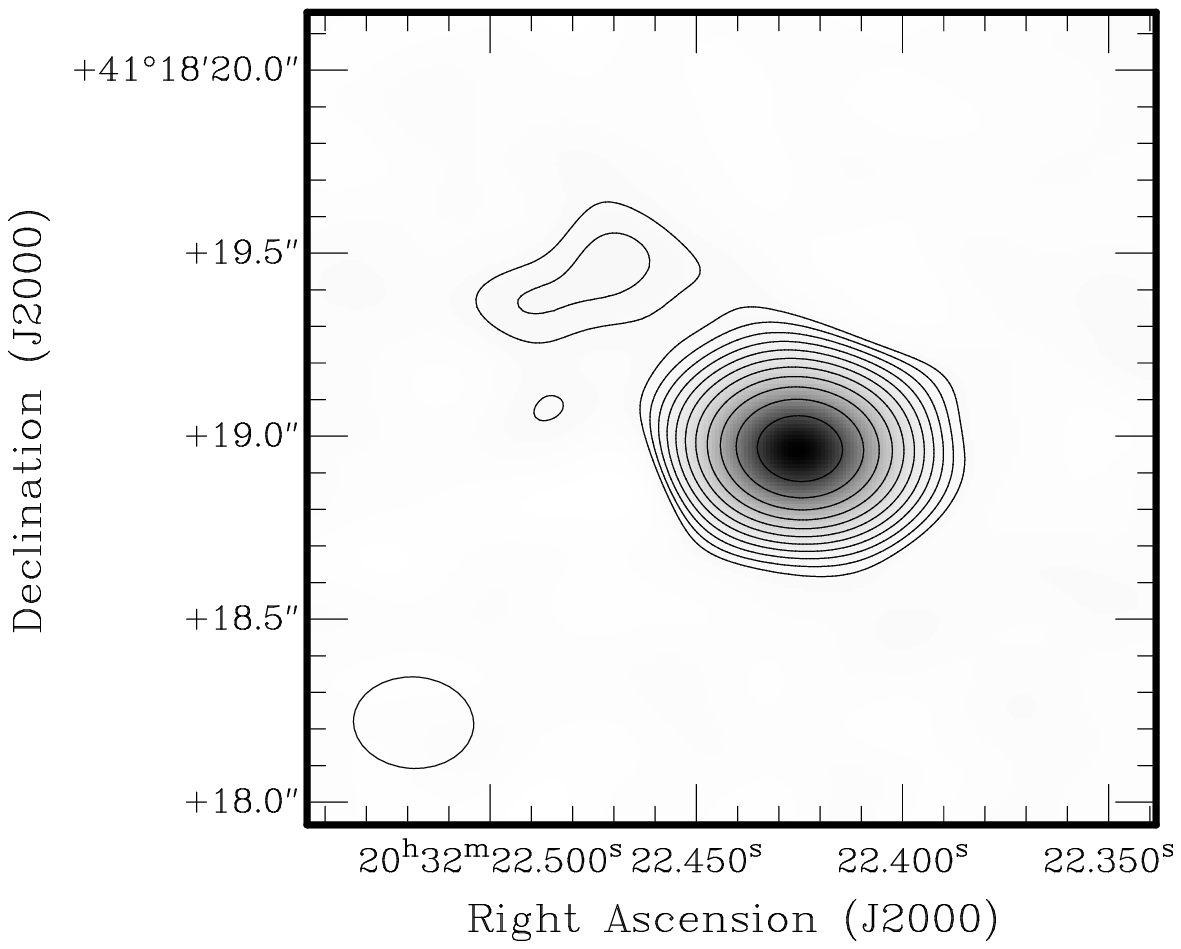}
\caption{ Two examples of the deconvolved VLA images at 8.4~GHz that
  show the primary and NE sources. The left image is from 1992
  December 19 during a low emission state and the right image is from
  1996 December 28 during a high emission state. \label{fig:images}}
\end{figure}

At 8.4 GHz, the two components were readily resolved in all
observations obtained with A and B configuration of the VLA, whereas
at 4.8 GHz the two components were only resolved in A-configuration
observations. In all of these observations, NE was always detected. In
C and D configurations, the two components are not resolved separately
at any of the observing frequencies, with only a single unresolved
source being observed. Higher resolution MERLIN observations at 1.4
GHz resolve the two radio components separately giving a 1.4-GHz flux
for NE and confirming its non-thermal nature with a negative spectral
index..

\section{Analysis: variations in Radio Emission}

The fluxes of {Cyg OB2 \#5} at both 4.8~GHz and 8.4~GHz as a function
of time are shown in Fig.~\ref{fig:light_curves}. Through the 20 years
of observations it is evident that the 4.8-GHz emission from {Cyg OB2
  \#5} has cycled through three cycles of high and low emission with
an approximate period of $\sim7$~yrs. It is also clear that the radio
emission from NE shows no variation and the primary radio component is
the source of the variations in {Cyg OB2 \#5}.

\begin{figure}[h]
\includegraphics[width=8cm]{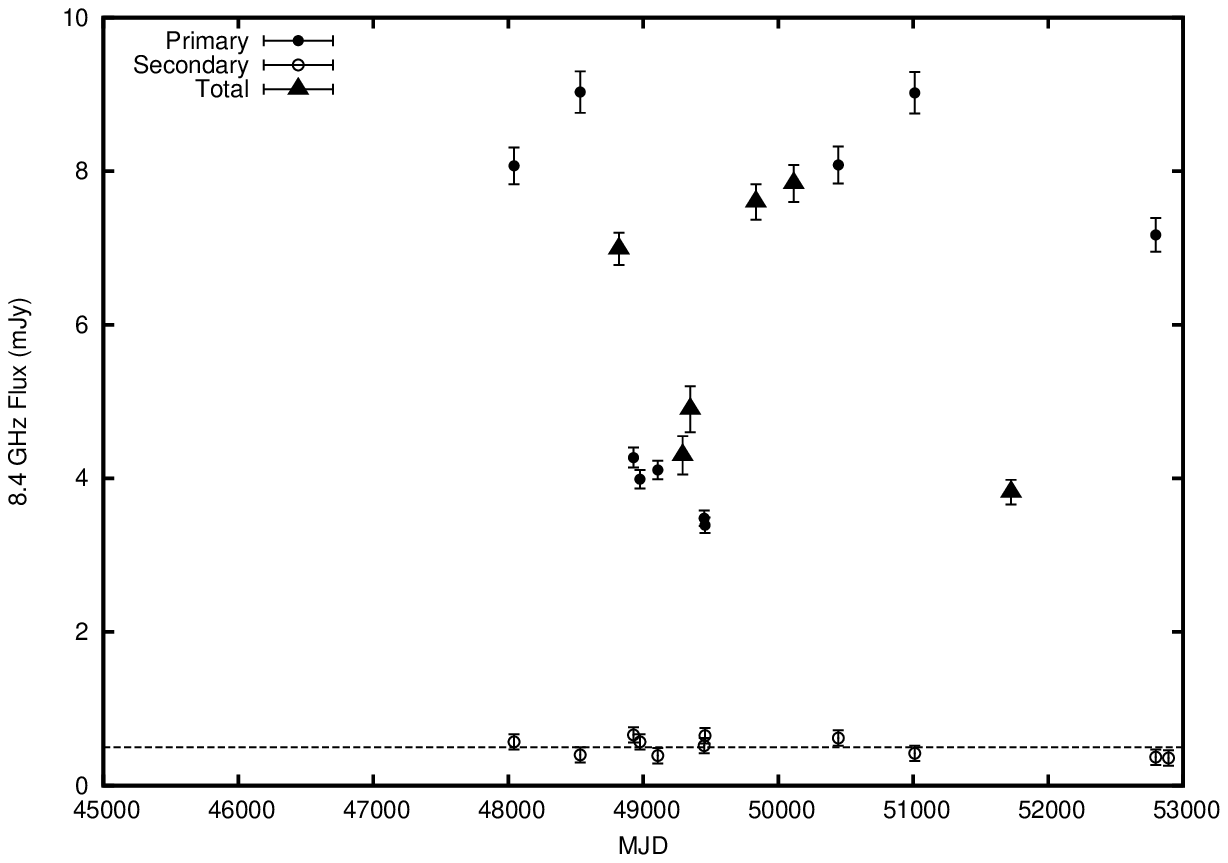}
\includegraphics[width=8cm]{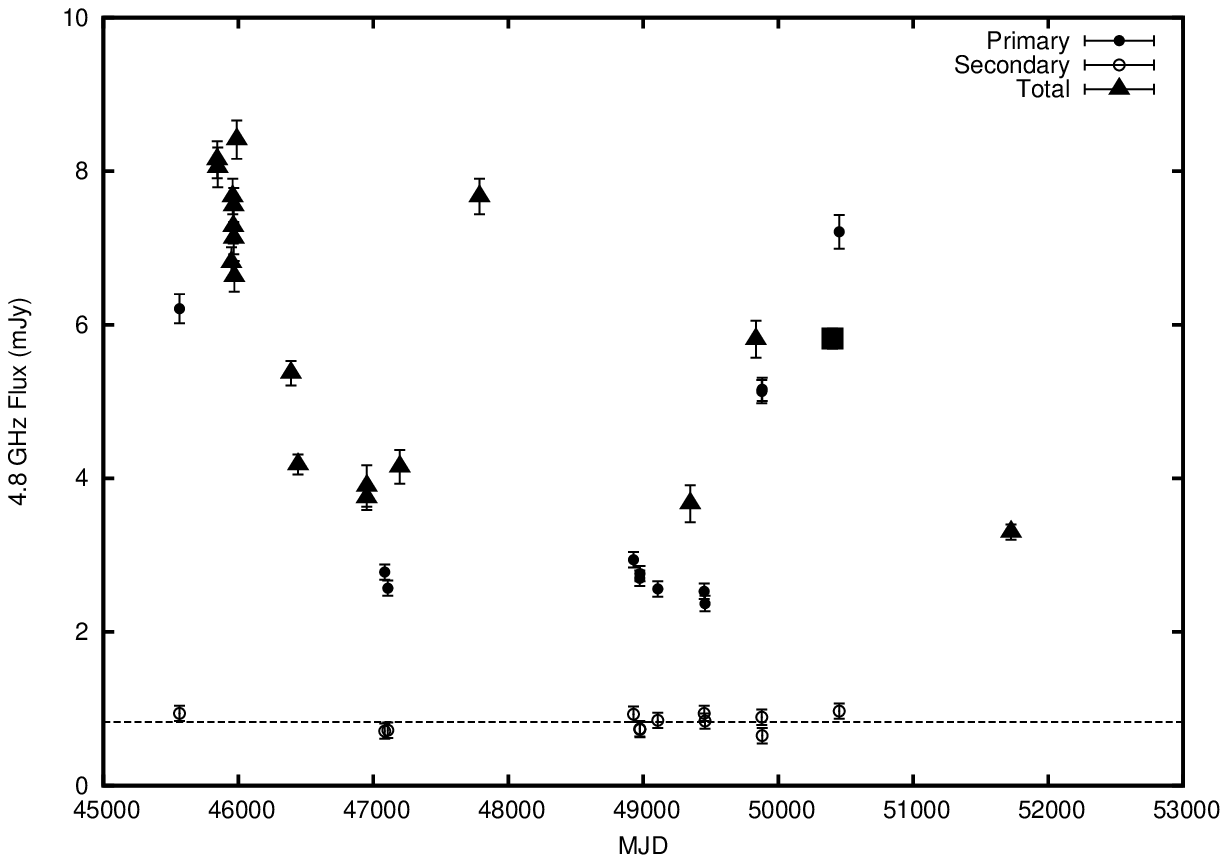}
\caption{ The fluxes of Cyg OB2 \#5 between 1983 and 2003 at 8.4~GHz
  (left) and 4.8~GHz (right), with the primary (solid circles) and NE
  (open circles) and total i.e.~when the components were {\em not}
  resolved as separate sources (solid triangles).  The mean fluxes of
  NE are shown (dashed line). The MERLIN 4.8-GHz observation is shown
  as a square. \label{fig:light_curves}}
\end{figure}

At both 4.8~GHz and 8.4~GHz there were 12 and 11 epochs of VLA
observations respectively where NE was resolved separately from the
primary and a flux could be measured directly. These observations of
NE have a weighted-mean flux at 4.8 GHz of $0.83\pm0.11$~mJy and
$0.50\pm0.12$~mJy at 8.4 GHz. The flux of NE is statistically
constant. This implies the variations arise exclusively in the primary
radio source.

Knowing that NE is constant enables the flux of the primary to be
determined for all observations where the two radio components are not
spatially resolved.  The period of the flux variations in the primary
emission was determined using a string-length technique (van Loo et
al. 2008, and references therein), and found to be $6.7\pm0.3$ years,
remarkably similar to the ``eyeball'' period derived in previous
works.

\section{Discussion}
\subsection{The Primary and Variable Emission}
\label{sec:primary_emission}
The primary emission component is associated with the O-star binary
system and is the source of all the observed variations in the radio
emission. In the low state the primary radio emission is found to have
a spectral index of $0.60\pm0.04$, consistent with that expected for
thermal emission arising in a steady-state radially symmetric stellar
wind. The thermal emission must be reasonably constant since 350-GHz
JCMT observations obtained away from radio minimum are
consistent with the stellar wind spectrum deduced from the radio
observations at radio minimum, with a best-fit spectral index
$0.63\pm0.04$ across this broad frequency range.  

During the high state, the spectral index is much flatter, with an
index of $0.24\pm\pm0.01$.  A model for the primary radio emission
component is proposed, where the lower spectral index during high
emission is the result of the addition of a non-thermal component to
the thermal emission from the binary system that leads to a ``composite''
spectrum. Such a model has been successfully applied to describe the
relatively flat continuum spectra of some Wolf-Rayet stars
(e.g. Chapman et al., 1999) where the non-thermal emission
arises in a WCR between the wind of the WR star and that of a massive
companion star.

For a non-thermal source embedded in a thermal stellar wind plasma, the total
observed flux as a function of frequency $\nu$ and at epoch $t$ is the sum
of the thermal and non-thermal fluxes, given by:
\begin{displaymath}
S_{obs}(\nu,t) = 2.5 \left(\frac{\nu}{4.8}\right)^{0.6} + S_{4.8}(t)
\left(\frac{\nu}{4.8}\right)^{\alpha} e^{-\tau(\nu,t)}~~~~~{\rm mJy},
\label{eqn:obs_flux}
\end{displaymath}
where it is assumed the thermal emission component is constant with a
spectral index of $+0.6$ and 4.8~GHz flux of 2.5~mJy measured in the
low, thermal emission state, and $S_{4.8}(t)$ is the intrinsic 4.8-GHz
flux of the non-thermal source at epoch $t$, $\alpha$ is the spectral
index of the intrinsic non-thermal emission assumed to be
constant. $\tau(\nu,t)$ is the line-of-sight free-free opacity through
the stellar wind to the non-thermal source at frequency $\nu$ and
epoch $t$, approximated by
\begin{displaymath}
\tau(\nu,t) \approx \tau_{4.8}(t)\left(\frac{\nu}{4.8}\right)^{-2.1}
\end{displaymath}
where $\tau_{4.8}(t)$ is the 4.8-GHz line-of-sight free-free opacity
at epoch $t$. The line-of-sight opacity is dependent on the geometry of the
line-of-sight to the non-thermal emission. Here, the case of a
non-thermal source in a 6.7-yr orbit about the binary is
considered. Williams et al. (1990) derived the varying free-free
opacity along a line-of-sight to a non-thermal source orbiting in the
circumbinary wind of the massive WR+O binary WR\,140.  The opacity is
dependent on the orbit inclination ($i$), argument of periastron
($\omega$), as well as the eccentric anomaly ($e$) and the time of
periastron passage ($T_{0}$), hence
\begin{displaymath}
\label{eqn:williams}
\tau_{4.8}(t) = \tau_{4.8}(t,i,\omega,e,T_{0})
\end{displaymath}
The intrinsic non-thermal flux $S_{4.8}(t)$ is expected to depend on
the local conditions e.g. electron density, which will vary as the
source moves through the dense circumbinary wind. This may be
approximated by assuming a simple power-law relation with separation,
namely
\begin{displaymath}
S_{4.8}(t) = S'_{4.8} r^{-s},
\label{eqn:s48propto}
\end{displaymath}
where $S'_{4.8}$ is the non-thermal flux when the separation is equal
to $a$, and $s$ is a power-law index.  These definitions, along with
the analytic form for $\tau_{4.8}(t)$ cf. Williams et al. 1990, allow
$S_{obs}(\nu,t)$ to be determined as a function of the orbital phase
of the non-thermal source orbiting the binary system. The resulting
light curves arising from these models are plotted in
Fig.~\ref{fig:spectrum_model} for four different model parameter sets,
showing excellent agreement with the observations.

\begin{figure}[h]
\includegraphics[width=8cm]{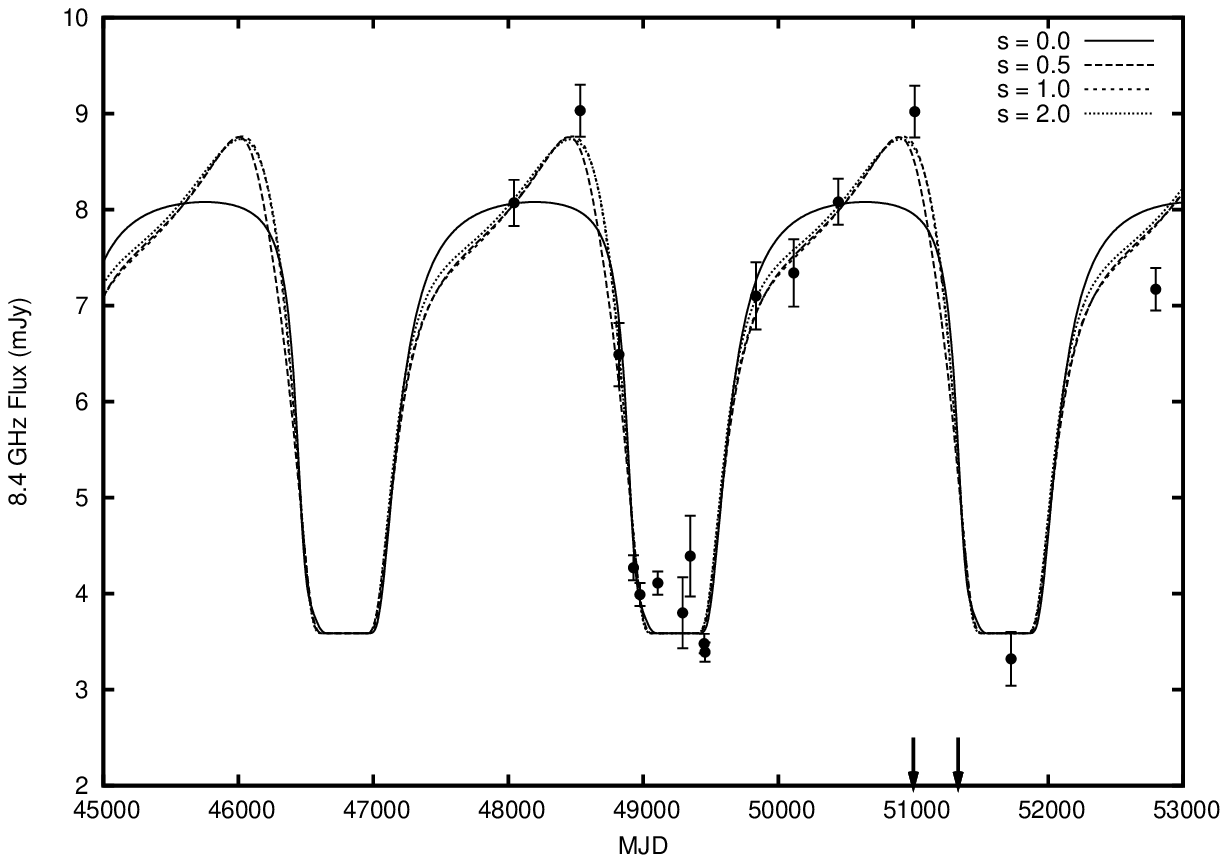}
\includegraphics[width=8cm]{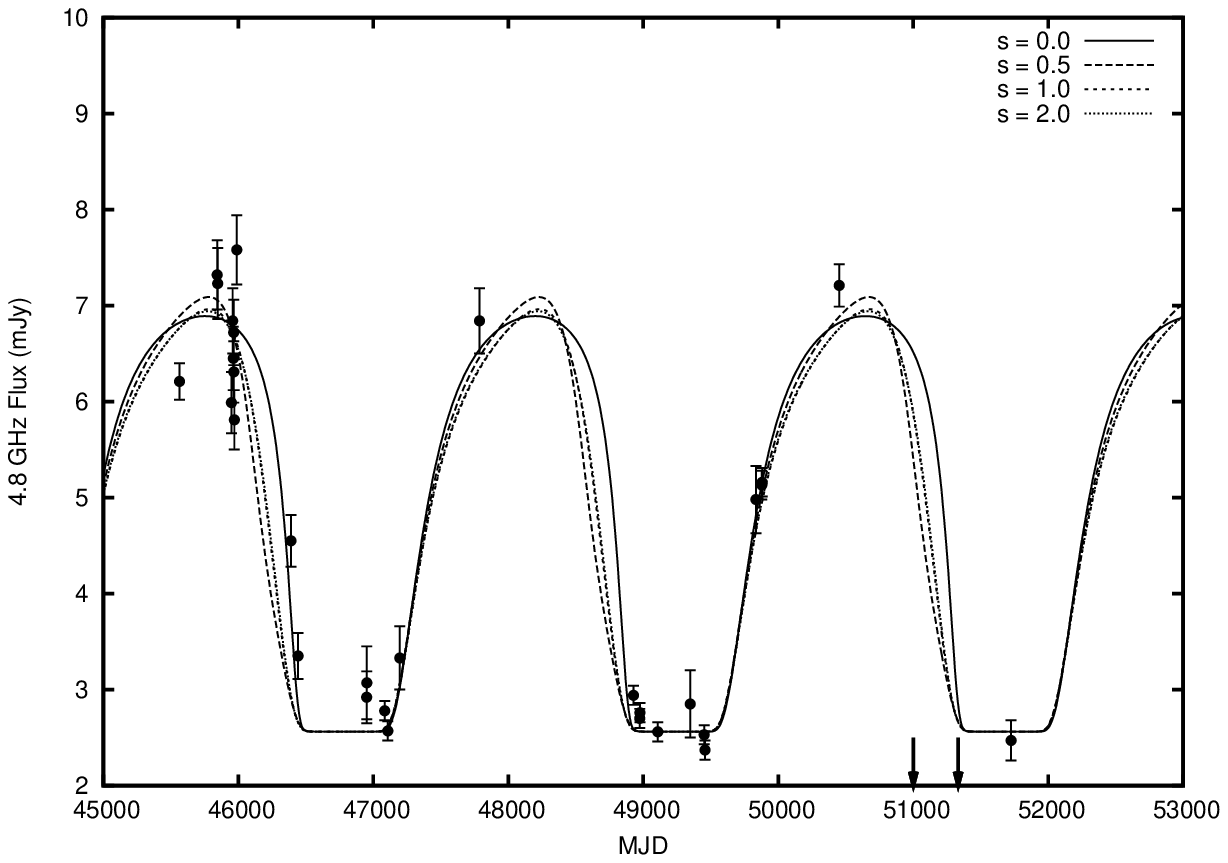}
\caption{ The best-fit orbiting non-thermal source model against the
  observed fluxes of the primary (slid circles) at 8.4~GHz (left) and
  4.8~GHz (right) for the $s=0, 0.5, 1$, and 2
  models.\label{fig:spectrum_model}}
\end{figure}

\subsection{Evidence for a Third Star?}
\label{sec:thirdstar}
A non-thermal source orbiting the O+O binary system requires a star
(hereafter Star C) to be in a 6.7-year orbit around the binary.  This
star may contribute the non-thermal radio emission via a WCR arising
from the collision of its own stellar wind with the wind from the O+O
star binary. Such WCRs have been
observed directly in some WR+O star and O+O star binary systems
(e.g. Dougherty \& Pittard, 2006, and references therein).
Alternatively, the non-thermal emission may arise from the putative
third star directly, e.g. a compact object.

Given the high luminosities of the two supergiants in the binary and
emission from circumstellar material, it will be very hard to detect
the proposed third star directly, let alone measure its orbit.
Instead, the radial velocities (RVs) of the O+O binary from Rauw
et. al. (1999) and Bohannan \& Conti (1976) are examined to search for
reflex motion due to its putative orbit with Star C.
For each RV observed from the
primary, the residual (O--C) was calculated from the orbit by Rauw
et. al. (1999) and formed the average (O--C) for each run.  A
systematic increase of RV between phases 0.26 and 0.89 is seen,
implying that the O+O binary moves away from us more rapidly.  This
implies that Star C moves towards us more rapidly in this phase
interval so that the circumbinary extinction to the non-thermal radio
source diminishes, consistent with it brightening during this orbital
phase.

The run of mean (O--C) with phase is compared with the reflex motion
of the O+O binary in orbit with Star C following the orbital elements
of the embedded non-thermal radio source from the $s=0$ case (see
Fig.~\ref{fig:RVOmC}). Fitting
\begin{displaymath}
{\rm v}_{r} = \gamma+K_{{\rm O+O}}\bigl( e \cos\omega + \cos(f + \omega) \bigr)
\end{displaymath}
for $K_{{\rm O+O}}$ and systemic velocity $\gamma$, gives $K_{{\rm
O+O}} = 32\pm17$~km\,s$^{-1}$ and
$\gamma=-5.9\pm4.7$~km\,s$^{-1}$. This leads to a mass function $f(m)$
derived from $P$ (in days) and $K$ (in km\,s$^{-1}$) from

\begin{displaymath}
f(m) = \frac{m^3_{\rm C}\sin^3(i)}{(m_{{\rm O+O}}+m_{\rm C})^2} =
1.036 \times 10^{-7} \left(1 - e^{2}\right)^{3/2} K^{3} P,
\end{displaymath}
giving an estimate of the mass, $m_{\rm C}$, of Star C. From the
data here, $f(m)=3.2^{+8.2}_{-2.8}~{\rm M}_{\odot}$. Assuming
$\sin(i)=1$ and adopting $m_{{\rm O+O}} = 41.5\pm3.4$~M$_{\odot}$ from
Linder et. al. (2009), this gives $m_{\rm C} =
23^{+22}_{-14}$~M$_{\odot}$ for Star C.

\begin{figure}[h]
\centering
\includegraphics[width=8cm]{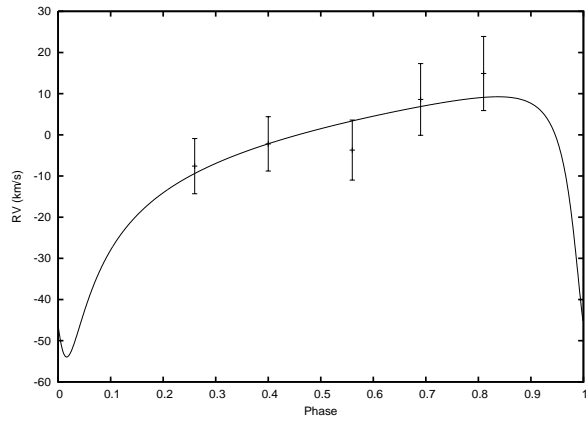}
\caption{ Observed (O--C) residuals and the RV curve for the reflex
  motion for the elements of the $s=0$ model with
  $K=32\pm17$~km\,s$^{-1}$ and
  $\gamma=-5.9\pm4.7$~km\,s$^{-1}$. \label{fig:RVOmC}}
\end{figure}

%
%
 
%
%
\footnotesize
\beginrefer
\refer Abbott, D. C., Bieging, J. H., \& Churchwell, E. 1981, ApJ, 250, 645

\refer Bieging, J. H., Abbott, D. C., \& Churchwell, E. B. 1989, ApJ, 340, 518

\refer Bohannan, B., \& Conti, P. S. 1976, ApJ, 204, 797

\refer Chapman, J. M., Leitherer, C., Koribalski, B., Bouter, R., \& Storey, M. 1999, ApJ, 518, 890

\refer Contreras, M. E., Rodriguez, L. F., Tapia, M., Cardini, D., Emanuele, A., Badiali, M., \& Persi, P. 1997, ApJl, 488, L153

\refer Dougherty, S., \& Pittard, J. M. 2006, in Proceedings of the 8th European VLBI Network Symposium

\refer Hall, D. S. 1974, Acta Astronomica, 24, 69

\refer Kennedy, M., Dougherty, S.M., Williams, P.M., Fink, A. 2010, ApJ, 709, 632

\refer Leung, K.-C., \& Schneider, D. P. 1978, ApJ, 224, 565

\refer Linder, N., Rauw, G., Manfroid, J., Damerdji, Y., De Becker, M., Eenens, P., Royer, P., \& Vreux, J.-M. 2009, A\&A, 495, 231

\refer Miralles, M. P., Rodriquez, L. F., Tapia, M., Roth, M., Persi, P., Ferrari-Toniolo, M., \& Curiel, S. 1994, A\&A, 282, 547

\refer Persi, P., Ferrari-Toniolo, M., \& Grasdalen, G. L. 1983, ApJ, 269, 625

\refer Persi, P., Ferrari-Toniolo, M., Tapia, M., Rodriguez, L. F., \& Roth, M. 1990, A\&A, 240, 93

\refer Rauw, G., Vreux, J.-M., \& Bohannan, B. 1999, ApJ, 517, 416

\refer van Loo, S., Blomme, R., Dougherty, S. M., \& Runacres, M. C. 2008, A\&A, 483, 585

\refer Williams, P. M., van der Hucht, K. A., Pollock, A. M. T., Florkowski, D. R., van der Woerd, H., \& Wamsteker, W. M. 1990, MNRAS, 243, 662

\endrefer
     
\end{document}